\documentclass[]{article}
\usepackage{amsfonts}

\title{Entanglement of a microcanonical ensemble}
\author{Tobias Verhulst \\
\footnotesize\it
 Departement Fysica, Universiteit Antwerpen, Groenenborgerlaan 171, 2020 Antwerpen,
 Belgium,\\
\footnotesize tobias.verhulst@ua.ac.be\\
[2ex]
        Jan Naudts      \\
\footnotesize\it
 Departement Fysica, Universiteit Antwerpen, Groenenborgerlaan 171, 2020 Antwerpen,
 Belgium,\\
\footnotesize jan.naudts@ua.ac.be
} 
\date {}

\newcommand{\be}{\begin{eqnarray}}
\newcommand{\ee}{\end{eqnarray}}

\newcommand{\TrA}{\,{\rm Tr_{_A}}\,}
\newcommand{\TrB}{\,{\rm Tr_{_B}}\,}
\newcommand{\EA}{{\cal E}_{_A}}
\newcommand{\EB}{{\cal E}_{_B}}

\newcommand{\Io}{{\mathbb I}}

\newcommand{\rhoA}{\rho_{_A}}
\newcommand{\rhoB}{\rho_{_B}}

\begin{document}

\maketitle
\begin{abstract}
We replace time-averaged entanglement by ensemble-averaged entanglement
and derive a simple expression for the latter. We show how to calculate
the ensemble average for a two-spin system and for the Jaynes-Cummings
model. In both cases the time-dependent entanglement is known as well
so that one can verify that the time average coincides with the ensemble
average.
\end{abstract}

\section{Introduction}

The entanglement of particles is in principle a time-dependent
quantity. This time-dependence has been analysed recently in
chaotic systems \cite {WGSH04}, in experimental spectra
of triatomic molecules \cite {HWM06}, and in Rydberg atoms \cite{LM06}.
Time-dependent entanglement has been studied in theoretical models,
like the Dicke model \cite {HH04}, a model of coupled kicked tops
\cite {DDK04}, the Harper Hamiltonian \cite {LS05}, a dimer model \cite {HCH05},
Bose-Einstein condensates \cite {XH05}.
In these papers, the notions of time-averaged entanglement and
ensemble-averaged entanglement
have been shown to be useful in monitoring phase transitions,
although the generality of this relationship has been questioned, see e.g.~\cite{LM06}.

In addition, time-averaged and ensemble-averaged entanglement are
conserved quantities of  quantum microcanonical ensembles \cite {NVdS06}.
As such they are of interest in the study of closed systems.
This context is suited to discuss the relation between both concepts,
and is the starting point of the present paper.

The entanglement of formation of a pure state is taken here
to be defined as the von Neumann
entropy of the reduced density matrix. Often, the von Neumann entropy
is replaced by the linear entropy because the latter can be computed more easily.
In the present paper the use of the linear entropy is essential
to obtain simple results.

For the sake of completeness, and to fix notations,
the definitions of entanglement of
pure and of mixed states are reproduced in the next section.
Section 3 introduces the entanglement of microcanonical ensembles
of wavefunctions.
The main result is announced in Section 4. The proof
is found in the Appendix.
It is followed by a section devoted to the complications
that arise when the state of the system has additional symmetries.
In Sections 6 and 7 the main result is applied to a system
of two interacting spins. Section 8 deals with the
Jaynes-Cummings model. For this model the time-dependent
entanglement is known so that it can be compared with the ensemble
average. A short discussion follows in Section 9.

\section{Definition of entanglement}

Consider two independent subsystems, labelled A and B.
With each normalised wavefunction $\psi$ of the combined system corresponds
a reduced density matrix $\rhoA$ of the subsystem labelled $A$.
The latter is defined by the relation
\be
\TrA\rhoA X=\langle\psi|X\otimes\Io|\psi\rangle
\qquad\mbox{ for all }X.
\ee
The entanglement of $\psi$ is then equal to the von Neumann
entropy of $\rho_A$. For technical reasons we replace this
entropy by the linear entropy.
The general definition of entanglement is
\be
\EA(\psi)=S_f(\rhoA)\equiv \TrA \rhoA f(\rhoA),
\ee
with $f(x)=-\ln x$ in the von Neumann case, and $f(x)=1-x$ in the case
of the linear entropy.

If $\psi$ is of the product form $\psi=\psi_{_A}\otimes \psi_{_B}$
then $\rhoA$ is a one-dimensional
projection operator. Hence, the entanglement vanishes.
A similar definition holds for $\rhoB$ and for $\EB$.
The entanglements $\EA(\psi)$ and $\EB(\psi)$ are equal \cite {BBPS96}.
To see this, select a basis $u_m$ in subsystem A and a basis
$v_p$ in subsystem B, so that
\be
\psi=\sum_n\sqrt {p_n}u_n\otimes v_n.
\ee
This is possible by means of the Schmidt construction.
Then $\rhoA$ and $\rhoB$ are diagonal, with eigenvalues $p_n$,
and with entropy equal to $\sum_np_nf(p_n)$.

Often, the state of the system is not described by a wavefunction $\psi$
but by a density matrix $\rho$. Such a density matrix can be written
into the form
\be
\rho=\sum_np_n|\psi_n\rangle\,\langle\psi_n|,
\label {rhodecomp}
\ee
with $p_n\ge 0$, $\sum_np_n=1$, and with $\psi_n$
normalised wavefunctions.
Then the entanglement of $\rho$ has been defined \cite {BDSW96}
as the minimum of the average entanglement
\be
{\cal E}(\rho)=\min\sum_np_n\EA(\psi_n),
\label {wootters}
\ee
where the minimum is taken over all possible ways to write (\ref {rhodecomp}).

\section {Definition of mean entanglement of a microcanonical ensemble}

The mean entanglement, which
is studied in the present paper,  is {\sl not} the average (\ref {wootters}),
but rather the average over a microcanonical ensemble, as introduced
in \cite {NVdS06}.

Let be given a density matrix $\rho$, which is diagonal in the orthonormal
basis of wavefunctions $\psi_n$, with eigenvalues $p_n$: $\rho\psi_n=p_n\psi_n$.
Associated with this diagonal density matrix is an ensemble of wavefunctions
of the form
\be
\psi=\sum_n\sqrt {p_n}e^{i\chi_n}\psi_n,
\label {ensemble}
\ee
where the $\chi_n$ are arbitrary phase factors.
The ensemble average of the entanglement is then denoted
$\overline {{\cal E}}$ and is given by
\be
\overline {{\cal E}}=
\left\langle\EA\left(\sum_n\sqrt {p_n}e^{i\chi_n}\psi_n\right)
\right\rangle_\chi,
\label {meanent}
\ee
where the average over $\chi$ is obtained by integrating over all phase factors
$\chi_n$ from 0 to $2\pi$, normalised by dividing by a factor $2\pi$.
Note that the ensemble average (\ref {meanent}) does not depend on the
chosen subsystem because $\EA(\psi)=\EB(\psi)$ for all $\psi$.

The ensemble (\ref {ensemble}) can be obtained by starting from a single
wavefunction $\psi$, in combination with the quantum mechanical time evolution.
The Hamiltonian $H$ is the generator of the unitary time evolution
\be
\psi_t=U(t)\psi
\qquad\mbox{ with }\qquad
U(t)=e^{-i\hbar^{-1}tH}.
\ee
The time average of the entanglement $\EA(\psi)$ is then defined by
\be
\left\langle \EA(\psi_t)\right\rangle_t
=\lim_{T\rightarrow\infty}\frac 1T\int_0^T{\rm d}t\,\EA(\psi_t).
\ee
Assume now that the Hamiltonian is diagonal in the basis of
wavefunctions $\psi_n$, with eigenvalues $\epsilon_n$.
Then one has
\be
\psi_t=\sum_n\lambda_ne^{-i\hbar^{-1}\epsilon_nt}\psi_n
\label {tewf}
\ee
and hence
\be
\left\langle \EA(\psi_t)\right\rangle_t
=\lim_{T\rightarrow\infty}\frac 1T\int_0^T{\rm d}t\,
\EA\left(\sum_n\lambda_ne^{-i\hbar^{-1}\epsilon_nt}\psi_n
\right).
\label {tav}
\ee
The wavefunctions $\psi_t$ belong to the ensemble (\ref {ensemble})
with $p_n=|\lambda_n|^2$. If the conditions of
the classical ergodic theorem holds then the time average (\ref {tav})
coincides with the ensemble average (\ref {meanent}) --- see \cite {NVdS06}.
But even when the classical ergodic theorem does not hold one can
continue to use the ensemble average instead of the time average
because experimentally the slightest perturbation may restore ergodicity.

\section {Main result}

Let be given an ensemble of wavefunctions of the form (\ref {ensemble}).
With each of the basis vectors $\psi_n$ is associated a couple of reduced
density matrices $\rhoA$ and $\rhoB$. For convenience, these
will be denoted $\sigma_n$ and $\tau_n$. Introduce the density
matrices $\sigma$ and $\tau$, defined by
\be
\sigma=\sum_n p_n\sigma_n
\qquad\mbox{ and }\qquad
\tau=\sum_n p_n\tau_n.
\ee
In some sense, $\sigma$ is the ensemble average of $\rhoA$,
$\tau$ is the ensemble average of $\rhoB$.

Our main result is now that the mean entanglement, defined by (\ref {meanent}),
using the linear entropy $S_1$, can be written as
\be
\overline {{\cal E}}=S_1(\sigma)+S_1(\tau)-\Delta,
\label {main}
\ee
where $\Delta$ is a contribution common to both $S_1(\sigma)$ and $S_1(\tau)$.
It is given by
\be
\Delta=1-\sum_mp_m^2\TrA\sigma_m^2=1-\sum_mp_m^2\TrB\tau_m^2.
\ee
The proof of this relation is given in the Appendix.

The applications of (\ref {main}) are explored in later sections.

\section {Degeneracies}

The ensemble (\ref {ensemble}) is uniquely defined by the density operator $\rho$
in the case that the eigenvalues $p_n$ of $\rho$ are two-by-two distinct.
Then the eigenfunctions $\psi_n$ are unique up to a phase factor.
However, If some of the eigenvalues $p_n$ coincide then the the orthonormal basis
is non-unique. In particular, if a non-zero eigenvalue $p_n$ is
degenerate then different choices of orthonormal wavefunctions may
influence the value of $\overline {{\cal E}}$.
This shows that $\overline {{\cal E}}$ is the average entanglement of the
ensemble and is not suitable as a definition of the entanglement of $\rho$.

A similar question is whether the entanglement of the ensemble can be 
used as the definition of the mean entanglement of the wavefunction $\psi$.
Consider the situation that some of the eigenvalues $\epsilon_n$ of the
Hamiltonian $H$ are degenerate. Then the basis of wavefunctions $\psi_n$,
which diagonalises $H$, is
not uniquely defined (up to phase factors). In that case
the wavefunction $\psi$
should not be decomposed into an arbitrary diagonalising orthonormal basis.
Rather, it should be projected onto the invariant
subspaces of $H$. This determines in a unique way an orthonormal basis
which then can be used to form the ensemble associated with $\psi$. 
An example of the degenerate case follows below.

\section{Two-spin example}

The simplest example is that of two quantum spins, each described by Pauli
spin matrices, and a Hamiltonian $H$ which is diagonal in the basis
of wavefunctions
\be
\psi_1&=&\frac{1}{\sqrt{2}}(\left|\uparrow\uparrow\right\rangle
+\left|\downarrow\downarrow\right\rangle)\cr
\psi_2&=&\frac{1}{\sqrt{2}}(\left|\uparrow\uparrow\right\rangle
-\left|\downarrow\downarrow\right\rangle)\cr
\psi_3&=&\frac{1}{\sqrt{2}}(\left|\uparrow\downarrow\right\rangle
+\left|\downarrow\uparrow\right\rangle)\cr
\psi_4&=&\frac{1}{\sqrt{2}}(\left|\uparrow\downarrow\right\rangle
-\left|\downarrow\uparrow\right\rangle).
\ee
We assume that the energy levels are non-degenerate. Their actual value is not needed.

The reduced density matrices $\sigma_m$ and $\tau_m$, corresponding with $\psi_m$,
are all equal to $\frac 12\Io$. Hence, also the averages $\sigma$ and
$\tau$ are equal to $\frac 12\Io$. As a consequence, the linear entropies
$S_1(\sigma)$ and $S_1(\tau)$ both equal 1/2.
However, the common part $\Delta$ depends on the choice of wavefunction $\psi$.
One finds
\be
\Delta=1-\sum_mp_m^2\TrA\sigma_m^2=1-\frac 12\sum_mp_m^2.
\ee
The final result for the mean entanglement of $\psi$ is therefore
\be
\overline{\cal E}=\frac 12+\frac 12-\left(1-\frac 12\sum_mp_m^2\right)
=\frac 12\sum_mp_m^2.
\label {twospinment}
\ee
Note that this result lies between 1/4 and 1/2.

It is possible but tedious to verify by explicit calculation that
the mean entanglement (\ref {twospinment}) coincides with the
time average of $\EA(\psi_t)$, as it should be.

\section {Degenerate two-spin example}

Consider a two-spin system with energy
$-\epsilon$ for anti-parallel spins and $+\epsilon$ for parallel spins.
This is a degenerate limit of the previous example.
The ensemble, generated by an arbitrary $\psi$, now contains two free
phase factors instead of four. It
consists of all wavefunctions of the form
\be
e^{i\chi_1}\lambda_1\psi_1+e^{i\chi_2}\lambda_2\psi_2
\ee
with $\chi_1$ and $\chi_2$ arbitrary, and with
$\lambda_1\psi_1,\lambda_2\psi_2$ given by
\be
\lambda_1\psi_1&=&|\uparrow\downarrow\rangle\,\langle\uparrow\downarrow|\psi\rangle
+|\downarrow\uparrow\rangle\,\langle\downarrow\uparrow|\psi\rangle\cr
\lambda_2\psi_2&=&|\uparrow\uparrow\rangle\,\langle\uparrow\uparrow|\psi\rangle
+|\downarrow\downarrow\rangle\,\langle\downarrow\downarrow|\psi\rangle.
\ee
The coefficients $\lambda_1$ and $\lambda_2$ are chosen in such a way that
$\psi_1$ and $\psi_2$ are normalised. The reduced density matrices
are found to be
\be
p_1\sigma_1&=&\left(\begin {array}{lr}
p_{+-}
&0\cr
0
&p_{-+}
                          \end {array}\right)\cr
p_2\sigma_2&=&\left(\begin {array}{lr}
p_{++}
&0\cr
0
&p_{--}
                          \end {array}\right),
\ee
with $p_{+-}=|\langle\uparrow\downarrow|\psi\rangle|^2$
and similar definitions for $p_{-+}$, $p_{++}$ and $p_{--}$.
Similar expressions hold for $\tau_1$ and $\tau_2$
\be
p_1\tau_1&=&\left(\begin {array}{lr}
p_{-+}
&0\cr
0
&p_{+-}
                          \end {array}\right)\cr
p_2\tau_2&=&\left(\begin {array}{lr}
p_{++}
&0\cr
0
&p_{--}
                          \end {array}\right).
\ee

The mean entanglement can now be calculated using (\ref {main}).
The result is
\be
\overline{\cal E}=2p_{++}p_{--}+2p_{+-}p_{-+}.
\ee
In the notation of the previous section this becomes
\be
\overline{\cal E}=\frac 12(p_1-p_2)^2+\frac 12(p_3-p_4)^2,
\ee
which is less than (\ref {twospinment}) by the term $-p_1p_2-p_3p_4$.

\section{The Jaynes-Cummings model}

The Jaynes-Cummings model \cite {JC63,CF65} describes a two-level system interacting with
a harmonic oscillator. The latter represents a single mode of the electromagnetic
field in a cavity. The model has been studied extensively.

The Hamiltonian of the model reads
\be
H=\hbar\omega a^\dagger a
+\frac 12\hbar\omega_0\sigma_z+\hbar\kappa(a^\dagger\sigma_-+a\sigma_+),
\ee
with $a^\dagger$ and $a$ creation and annihilation operators of
the harmonic oscillator, and with the Pauli matrices $\sigma_z$
and $\sigma_\pm$ describing the two-level system.

Let $|g\rangle$ and $|e\rangle$ denote the ground state, respectively the excited state
of the two-level system. Let $|n\rangle$, $n=0,1,\cdots$ denote the eigenstates
of the harmonic oscillator. The eigenstates of the Jaynes-Cummings Hamiltonian
are explicitly known, see e.g.~\cite {RJC99}. An orthonormal basis of
eigenfunctions is given by
\be
\psi_0&=&|g\rangle\otimes|0\rangle\cr
\psi_{1,n}&=&\cos(\theta_n)|g\rangle\otimes |n+1\rangle
+\sin(\theta_n)|e\rangle\otimes|n\rangle\cr
\psi_{2,n}&=&-\sin(\theta_n)|g\rangle\otimes |n+1\rangle
+\cos(\theta_n)|e\rangle\otimes|n\rangle.
\ee
The angles $\theta_n$, $n=0,1,2,\cdots$ follow from the relation
\be
\tan\theta_n=\kappa\frac {\sqrt {n+1}}{\frac 12(\omega-\omega_0)+\lambda_n}
\ee
with
\be
\lambda_n=\sqrt{\frac 14(\omega-\omega_0)^2+\kappa^2(n+1)}.
\ee

The time-dependence of the reduced density matrix can be calculated explicitly if the
initial state is a product state with the two-level system in the excited state
and the harmonic oscillator is in the $n$-th eigenstate, see e.g.~\cite {RJC99}.
The result for the reduced state of the two-level system is
\be
\rhoA(t)=W_n(t) |g\rangle\,\langle g|+(1-W_n(t))|e\rangle\,\langle e|
\ee
with
\be
W_n(t)=2\gamma_n\sin^2\left(t\kappa\sqrt{n+1}\right)
\ee
and with
\be
\gamma_n=\frac 12\sin^2(2\theta_n).
\ee
The linear entanglement is therefore
\be
\EA(\psi_t)=1-W_n^2(t)-(1-W_n(t))^2.
\ee
The time average equals
\be
\left\langle \EA(\psi_t)\right\rangle_t=2\gamma_n-3\gamma_n^2.
\label {timav}
\ee

It is straightforward to calculate the reduced density matrices for
the eigenfunctions of the model. The result is
\be
\sigma_0
&=&\left|0\right\rangle\left\langle 0\right|,\\
\sigma_{1,n}
&=&\cos^2\theta_n\left|n+1\right\rangle\left\langle n+1\right|+\sin^2\theta_n\left|n\right\rangle\left\langle n\right|,\\
\sigma_{2,n}
&=&\sin^2\theta_n\left|n+1\right\rangle\left\langle n+1\right|+\cos^2\theta_n\left|n\right\rangle\left\langle n\right|,\\
\tau_0
&=&\left|g\right\rangle\left\langle g\right|,\\
\tau_{1,n}
&=&\cos^2\theta_n\left|g\right\rangle\left\langle g\right|+\sin^2\theta_n\left|e\right\rangle\left\langle e\right|,\\
\tau_{2,n}
&=&\sin^2\theta_n\left|g\right\rangle\left\langle g\right|+\cos^2\theta_n\left|e\right\rangle\left\langle e\right|.
\ee
Hence, it is straightforward to evaluate the mean entanglement $\overline{\cal E}$
for an arbitrary wavefunction $\psi$. However, we did not succeed
to rewrite the resulting expression in a simple and transparent way.

In the case that $\psi$ is of the product form
\be
\psi=|e\rangle\otimes|n\rangle=\sin(\theta_n)\psi_{1,n}+\cos(\theta_n)\psi_{2,n}
\ee
 one obtains
\be
\sigma&=&\sin^2(\theta_n)\sigma_{1,n}+\cos^2(\theta_n)\sigma_{2,n}\cr
&=&\gamma_n|n+1\rangle\,\langle n+1|
+\left(1-\gamma_n\right)|n\rangle\,\langle n|\cr
\tau&=&\sin^2(\theta_n)\tau_{1,n}+\cos^2(\theta_n)\tau_{2,n}\cr
&=&\gamma_n|g\rangle\,\langle g|
+\left(1-\gamma_n\right)|e\rangle\,\langle e|\cr
\Delta&=&1-\sin^4(\theta_n)\TrB\tau_{1,n}^2-\cos^4(\theta_n)\TrB\tau_{2,n}^2\cr
&=&1-(1-\gamma_n)^2.
\ee
This leads to the result
\be
\overline{\cal E}=2\gamma_n-3\gamma_n^2,
\ee
which is identical with the time-averaged result (\ref {timav}).

\section{Discussion}

The calculation of the time-dependence of the entanglement of a quantum system
is a hard problem. The average over time is more accessible because it
can be replaced by an ensemble average. This is in particular so when the entanglement
is defined using the linear entropy instead of the von Neumann entropy
of the reduced density matrix, because in that case there exists a simple expression
for the mean entanglement --- see (\ref {main}). We have used this expression
in a two-spin system and in the Jaynes-Cummings model. For these systems it
is feasible to calculate both the averages over time and over the ensemble
of wavefunctions. The results of the two calculations coincide, as it should be.

We have pointed out that a systematic degeneracy of the energy levels of
the Hamiltonian due to the presence of a symmetry influences the choice
of the ensemble of wavefunctions, used in the calculation of the average
entanglement. In the example of the two spin system the additional symmetry
leads to a reduction of the entanglement.

Finally let us note that the use of the linear entropy is rather essential
in our paper. It is of course possible as well to define the ensemble
average of the entanglement, based on the von Neumann entropy.
It is however unlikely that a simple formula like (\ref {main})
exists in that case.

\section* {Appendix 
}

Here, the proof of (\ref {main}) is given.
From the definition (\ref {meanent}) follows,
assuming a linear entropy,
\be
\overline {{\cal E}}
&=&1-
\left\langle\TrA\rhoA^2(\chi)
\right\rangle_\chi,
\ee
with $\rhoA(\chi)$ defined by
\be
\TrA\rhoA(\chi)X
&=&\left\langle \sum_m\sqrt {p_m}e^{i\chi_m}\psi_m\bigg|
X\otimes\Io\bigg|\sum_n\sqrt {p_n}e^{i\chi_n}\psi_n\right\rangle\cr
&=&\sum_{mn}\sqrt {p_mp_n}e^{-i(\chi_m-\chi_n)}\TrA\sigma_{mn}X
\ee
with $\sigma_{mn}$ defined by
\be
\TrA\sigma_{mn}X=\langle\psi_m|X\otimes\Io|\psi_n\rangle.
\ee
Hence, one obtains
\be
\overline {{\cal E}}
&=&1-
\left\langle
\TrA\left(\sum_{mn}\sqrt {p_mp_n}e^{-i(\chi_m-\chi_n)}\sigma_{mn}
\right)^2\right\rangle_\chi\cr
&=&1-\sum_{mn}\sum_{rs}\sqrt {p_mp_np_rp_s}
\left\langle e^{-i(\chi_m-\chi_n)}e^{-i(\chi_r-\chi_s)}\right\rangle_\chi
\TrA \sigma_{mn}\sigma_{rs}\cr
&=&1-\sum_{mn}p_mp_n\TrA\sigma_{mm}\sigma_{nn}
-\sum_{mn}p_mp_n\TrA\sigma_{mn}\sigma_{nm}\cr
& &+\sum_mp_m^2\TrA\sigma_{mm}^2.
\ee
Now use that $\sigma_{mm}\equiv\sigma_m$ to see that the
first two terms yield the contribution $S_1(\sigma)$.
The last term is $1-\Delta$. Hence, it remains to be shown that
\be
S_1(\tau)=1-\sum_{mn}p_mp_n\TrA\sigma_{mn}\sigma_{nm}.
\label {aremains}
\ee

Choose an orthonormal basis $u_r$ for the subsystem A
and an orthonormal basis $v_p$ for the subsystem B.
Then one has
\be
\sigma_{mn}=\sum_{rsp}\left[\langle\psi_m|u_s\otimes v_p\rangle\,
\langle u_r\otimes v_p|\psi_n\rangle\right]\,|r\rangle\,\langle s|
\ee
so that
\be
\TrA\sigma_{mn}\sigma_{nm}
&=&\sum_{rspq}\langle\psi_m|u_s\otimes v_p\rangle\,
\langle u_r\otimes v_p|\psi_n\rangle\cr
& &\times \langle\psi_n|u_r\otimes v_q\rangle\,
\langle u_s\otimes v_q|\psi_m\rangle\cr
&=&\sum_{pq}
\langle\psi_m|\left(\Io\otimes |v_p\rangle\,\langle v_q|\right)|\psi_m\rangle\,
\langle\psi_n|\left(\Io\otimes |v_q\rangle\,\langle v_p|\right)|\psi_n\rangle\cr
&=&\sum_{pq}\langle v_q|\tau_m|v_p\rangle\,\langle v_p|\tau_n|v_q\rangle\cr
&=&\TrB\tau_m\tau_n.
\ee
The relation (\ref {aremains}) now follows readily.

\section*{}

\end{document}